\documentclass[3p,times,twocolumn]{elsarticle}

\usepackage{ecrc}

\volume{00}

\firstpage{1}

\journalname{Nuclear Physics B Proceedings Supplement}

\runauth{}

\jid{nuphbp}

\jnltitlelogo{Nuclear Physics B Proceedings Supplement}

\usepackage{amssymb}
\usepackage[figuresright]{rotating}

\begin{document}

\begin{frontmatter}



\dochead{}

\title{Lepton number violation in tau lepton decays}


\author[cinv]{G. L\'opez Castro}
\ead{glopez@fis.cinvestav.mx}

\author[cinv]{N. Quintero}
\ead{nquintero@fis.cinvestav.mx}
\address[cinv]{Departamento de F\'\i sica, Cinvestav del IPN, Apartado Postal 14-740, 07000 M\'exico, D.F. M\'exico}

\begin{abstract}
Recent studies of novel four-body lepton number violating decays of $\tau$ leptons and neutral $B$  mesons are summarized and updated.  These decays are assumed to be enhanced by the exchange of resonant Majorana neutrinos. It is shown that the $\tau^- \to \pi^+l^-l^-\nu_{\tau}$ decay channels, with $l=e$ or $\mu$, can provide stronger constraints on the mixing vs. mass parameter space of resonant Majorana neutrinos than analogous three-body decays of charged $B$ mesons.   
\end{abstract}


\end{frontmatter}

% \linenumbers

\section{Introduction}
Total lepton number $L=L_e+L_{\mu}+L_{\tau}$ is an absolutely conserved quantum number in the Standard Model (SM). Some extensions of the SM include interactions that can induce $L$ non-conservation \cite{moha}. Minimal extensions of the SM aiming to include massive neutrinos can contain Majorana mass terms, like ${\cal L}_M= \overline{{\nu}_R^c}M_M \nu_R + {\rm h.c.}$, which provides an appealing mechanism that violates lepton number by two units ($\Delta L=2$) \cite{paul}. A clear signal of Majorana mass terms are $L$-number violating processes that involve the production of two equal-sign charged leptons, the most well known and widely studied example being neutrinoless double beta decay in nuclei \cite{rode}. 

In this contribution we consider the exchange of Majorana neutrinos as a source of $\Delta L=2$ lepton number violation (LNV) in decays of heavy flavors, and more specifically in four-body decays of the $\tau$ lepton. These Majorana neutrinos are assumed to be sterile, such that their coupling to the weak  charged current are very suppressed by tiny mixings with active neutrinos. Typical neutrino-exchange diagrams contributing to LNV in decays of the $\tau$ lepton are shown in Figure 1.

Under this scheme, the sensitivity of different heavy flavor LNV decays ($M$ denotes a vector or pseudoscalar meson and $l,l'=e, \mu, \tau$ whenever they are allowed by kinematics) 
\begin{eqnarray}
 D^+_{(s)}, B^+, B^+_c &\to& l^+l'^+M^- \nonumber \\
 D^0, B^0, B_s &\to& l^-l'^- M_1^-M_2^-\nonumber \\ 
 \tau^- &\to& l^+M_1^-M_2^- \nonumber \\
 \tau^- &\to& \nu_{\tau}l^-l'^-M^+ \nonumber 
\end{eqnarray}
is determined by comparing the mass scale of the exchanged Majorana neutrinos with typical energies of the decay process. Thus, we distinguish three cases \cite{atre}:
\begin{itemize}
\item If neutrinos are very light compared to their four-momenta in the propagator (actually, $m_{\nu}^2 << q^2$), the decay rates become sensitive to the {\it effective} Majorana mass defined by $\langle m_{ll'}\rangle\equiv \sum_i U_{li}U_{l'i}m_i$, where $U_{li}$ denote the mixings of light (active) neutrinos described by the PMNS matrix; 
\item  If neutrinos are heavy compared to the mass of the decaying state, the rate is sensitive to $\sum_N V_{lN}V_{l'N}/m_N$, where $V_{lN}$ are the mixings of light (active) and heavy (sterile) neutrinos of type $N$ (see definition in Section 2). 
\item Finally, if heavy neutrinos are of the order of the heavy flavor mass scale such that they can be produced on their mass-shell ($q^2=m_N^2$), the rates are largely enhanced due to the resonant effect associated to their decay widths $\Gamma_N$, with their decay amplitudes proportional to $\sum_{N}V_{lN}V_{l'N}/\Gamma_N$. This is the so-called {\it resonant enhancement mechanism} \cite{atre} for LNV decays and can occur only for time-like neutrino momenta as in the case of mesons and $\tau$ lepton decays .
\end{itemize}
Note that in the first two cases, the rates of  heavy flavor decays turn out to be very suppressed, making uninteresting their searches at flavor factories \cite{ali,atre1}. 
\begin{figure}
\vspace{-1.5cm}
\hspace{-1.5cm}
\includegraphics[width=11.5cm]{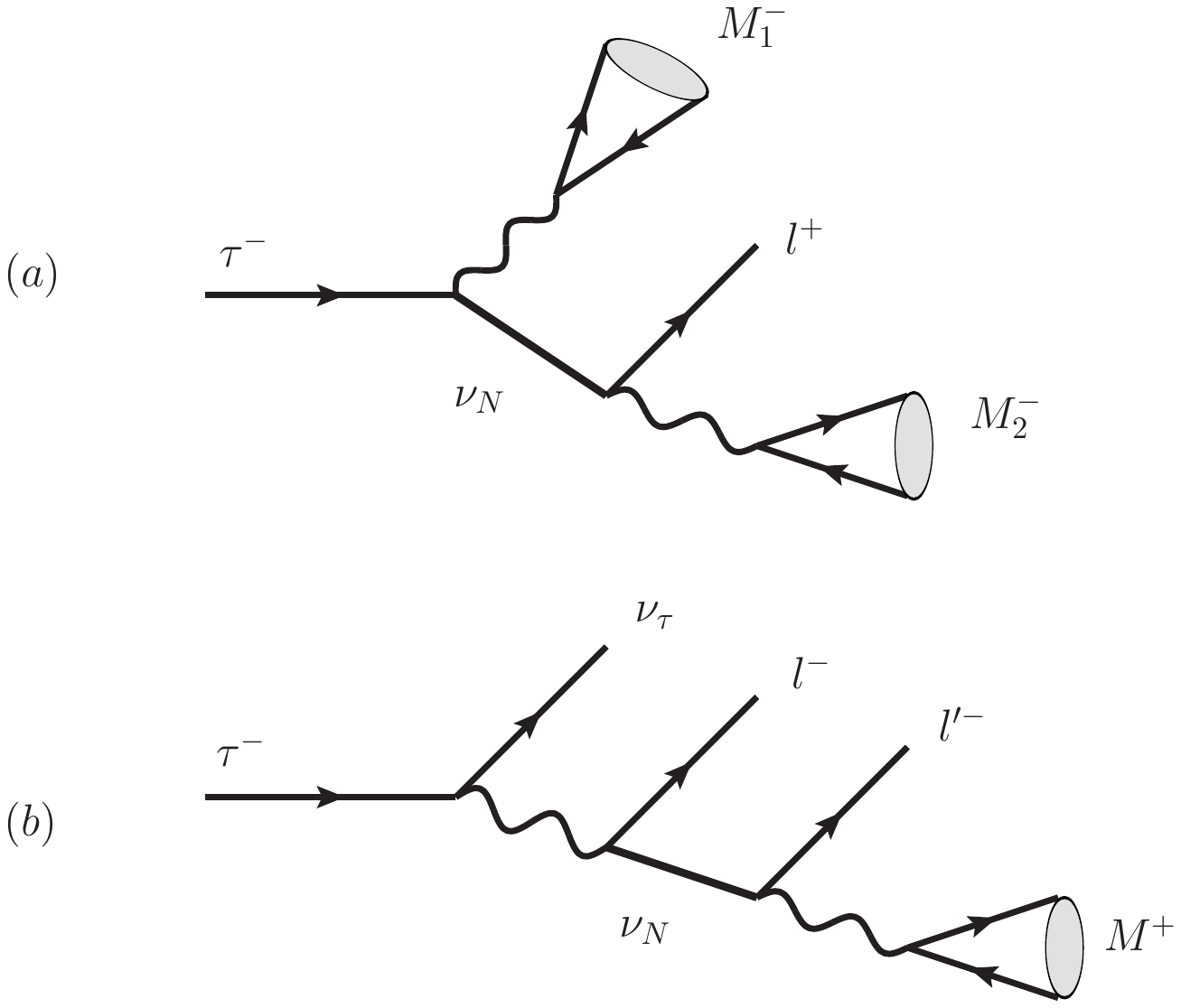}
\vspace{-9.0cm}
\caption{\small Neutrino-exchange diagrams induced by crossings of the $W^-W^-\to l^-l'^-$ $\Delta L=2$ kernel leading to LNV in $(a)$ three- and $(b)$ four-body tau decays.}
\end{figure}
Lepton number violation in three-body decays of $\tau$ leptons and charged ($D,\ D_s,\ B, \  B_c$)  mesons have been widely investigated previously, both from the theoretical \cite{atre,ali,atre1,chile1,bao} and experimental \cite{babar1, lhcb1, belle2, belle1, pdg, danielmstau2012, kiyoshitau2012} points of view. The current best experimental upper bounds available on these decay channels are shown in Figure 2; in addition, very stringent bounds of the order of $10^{-9}$ have been obtained (see for example \cite{pdg}) for $K^+ \to \pi^-l^+l'^+$ decays, with $l,\ l'=e,\ \mu$.  The measured upper limits allow to exclude a region in the $|V_{lN}|^2$ vs. $m_N$ plane of the parameter space, by assuming that a single resonant neutrino (usually denoted by the subindex $N$ or $4$) dominates the decay amplitude. Such sterile Majorana neutrinos, with masses in the range of 1$\sim$10 GeV, can appear in the framework of some minimal extensions of the SM; for instance, it has been suggested that they can play an important role to explain simultaneously the oscillations of neutrinos, the baryon asymmetry of the Universe and the dark matter problem \cite{shapo}. 

In this paper we present a summary and update of our recent proposals \cite{nos2011,nos2012} which consider the four-body decays $\tau^- \to \nu_{\tau}l^-l'^-M^+$ and $B^0 \to D^-l^+l'^+M^-$, where $M$ is a pseudoscalar or vector meson that can be allowed by kinematics (the analogous decay $\pi^+\to e^+e^+\mu^-\nu$ was considered in \cite{chile3}). We illustrate our studies with results on di-muonic channels (results on di-electrons modes can be found in \cite{nos2011,nos2012}). Searches for these decay channels have not been undertaken by experiments up to now. Here we emphasize that they can provide competitive or even stronger bounds on the parameter space of Majorana neutrinos as compared to three-body decays of heavy flavors.

\begin{figure}
\includegraphics[width=8.2cm]{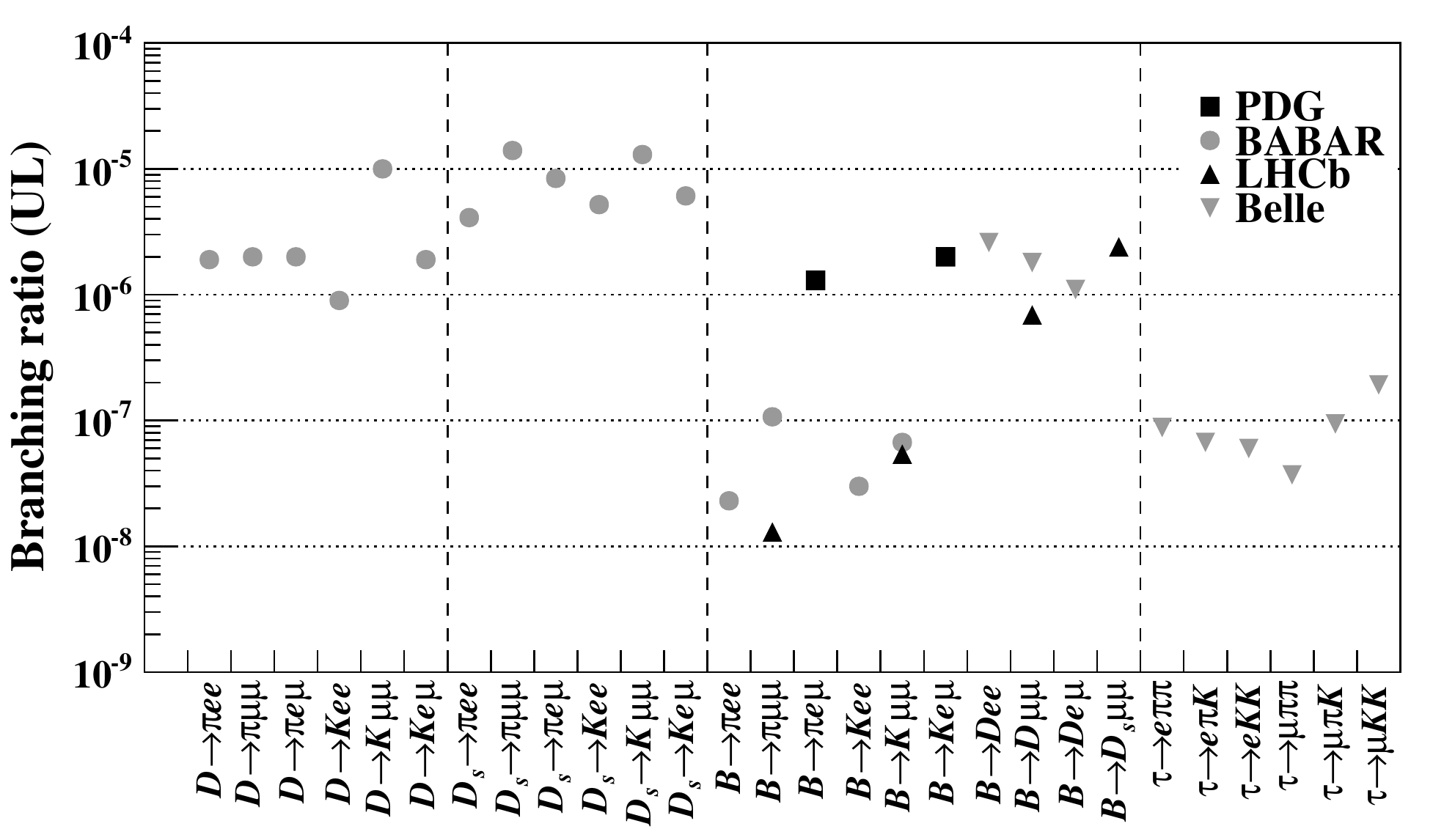}
\caption{\small Experimental upper limits on branching ratios of 3-body decays of charged heavy mesons and $\tau$ lepton [9-15]. }
\end{figure}

\section{Resonant three-body decays}

  The addition of right-handed singlet neutrinos to the SM leads in a natural way to the appearance of Majorana and Dirac mass terms \cite{paul}, with Majorana mass terms allowing $\Delta L=2$ lepton number violation. The heavier (sterile) neutrinos get involved into charged weak interactions, since after  diagonalization of the full neutrino matrix, neutrinos of defined flavor becomes a mixture of light and heavy mass eigentstates, namely, 
\begin{equation}
\nu_l=\sum_{i=1}^3 U_{li}\nu_i+\sum_{N= 4}^{n+4}V_{lN}\nu_N\ 
\end{equation} if $n$ right-handed singlets are considered. Here, $U_{li}$ are essentially the entries of the PMNS matrix, and $V_{lN}$ are the tiny mixings of the active and sterile neutrinos.
 The charged current interaction Lagrangian in the flavor basis becomes:
\begin{equation}
\!\!\!{\cal L}_{cc}=\frac{g}{2\sqrt{2}}\bar{\nu}_l\gamma^{\mu}(1-\gamma_5)l\cdot W^-_{\mu} + {\rm h.c.}
\end{equation}
where $\nu_l$ is given above.

Under the assumption that only one Majorana neutrino $N$ is resonant in three-body $\tau^- \to l'^+ M_1^- M_2^-$ and $M_1^- \to l^-l'^-M_2^+$ decays, the generic form of the decay amplitudes is (properly antisymmetrization under exchange of identical leptons in the final state must be understood)  
\begin{equation}
\!\!\!\!\!\! {\cal M}_{3}^{\tau,M_1^-}\sim G_F^2 V_{lN}V_{l'N}m_N{\cal F}(q^2)V_{M_1}^{CKM}V_{M_2}^{CKM}f_{M_1}f_{M_2}\ ,
\end{equation}
where $V^{CKM}_{M_i}$ is the Cabibbo-Kobayashi-Maskawa (CKM) matrix element for the charged meson $M_i$ and $f_{M_i}$ its decay constant. The resonance factor is determined by the neutrino propagator ${\cal F}(q^2)\sim (q^2-m_N^2+im_N\Gamma_N)^{-1}$ where $q$ is the momenta of the exchanged neutrino. The neutrino decay width $\Gamma_N\leq 10^{-3}$ eV for the mass scales of interest in $\tau$ lepton and in other me son decays ($m_N \leq 5$ GeV) \cite{atre}. 

From Eq. (3) we note that the bigger is the CKM matrix element, the stronger is the constraint that can be set on the neutrino mixings from the measured upper limits on LNV branching fractions. Since the rates  of $B^{\pm}$, $D^{\pm}$ meson decay vertices are Cabibbo-suppressed by $|V_{ub}|^2$ and $|V_{cd}|^2$ factors, respectively, the constraints that are derived from them will not be very strong . On another hand,  $\tau$ lepton decays (taking $l=\tau$ in Eq. (3) above) allow to constrain only the product $|V_{\tau N}V_{l'N}|$ ($l'=e,\mu$) from the measured upper limits of $\tau^- \to l'^+ M_1^-M_2^-$ branching fractions. 

 \begin{figure}
\includegraphics[width=8.5cm]{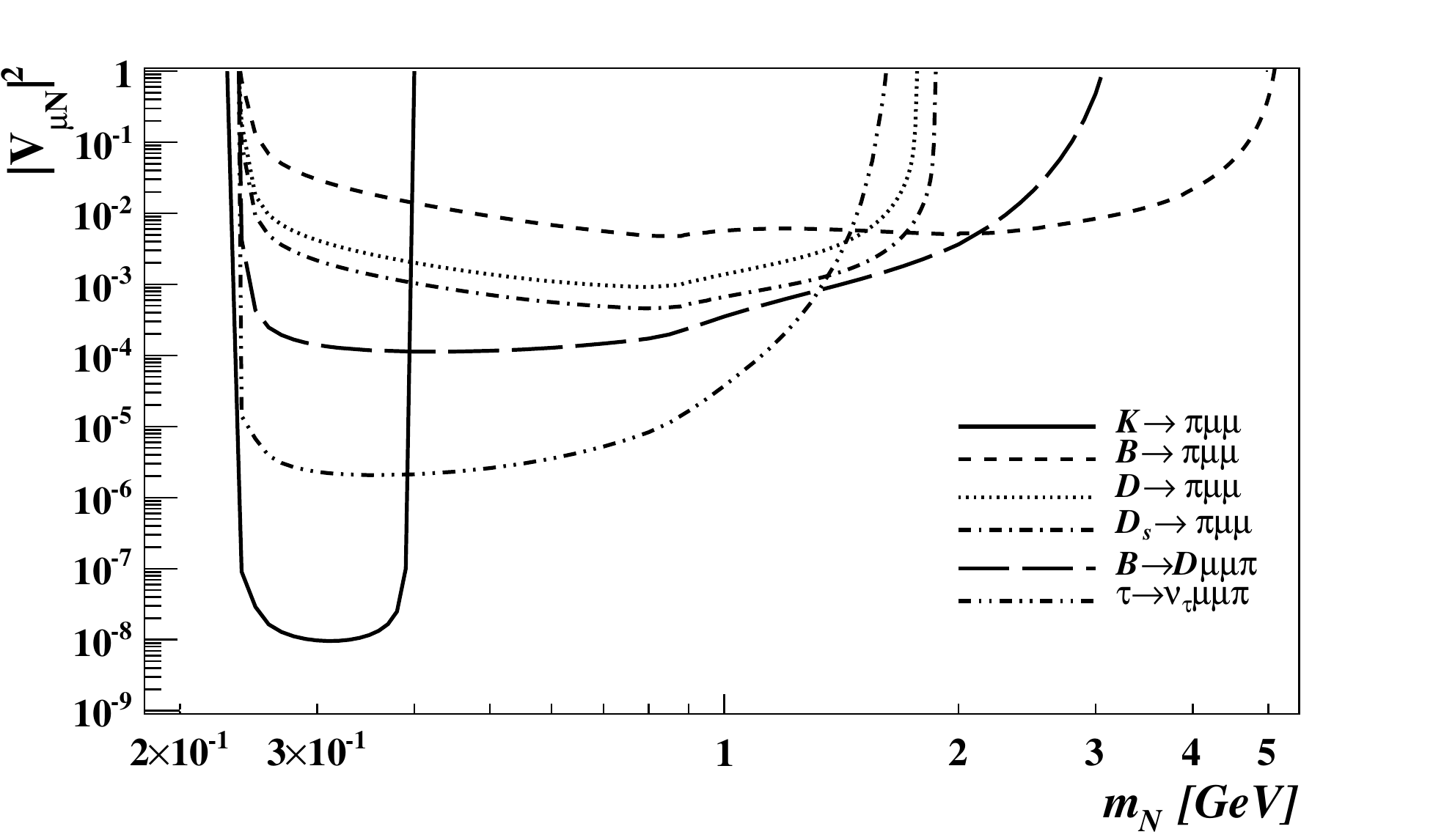}
\caption{\small Constrains on mixing vs. mass of Majorana neutrino parameter space from di-muonic three-body decays of charged mesons. Constraints from four-body $B^0 \to D^-\pi^-\mu^+\mu^+$ and $\tau^- \to \nu_{\tau}\mu^-\mu^-\pi^+$ by assuming upper bounds of $10^{-7}$ and $10^{-8}$ on their branching ratios, respectively, are also shown for comparison.}
\end{figure}
  In Figure 3 we show the  constraints in the $|V_{\mu N}|^2$ vs $m_N$ region that can be gotten from three-body LNV decays of charged $D_{(s)}$ and $B$ mesons by using the experimental upper limits on di-muonic channels. These plots were obtained by using the rates calculated in Ref. \cite{atre} and incorporate recently updated measurements obtained by $B$-factory experiments \cite{babar1} and LHCb \cite{lhcb1}. Note that despite the large improvement recently obtained by LHCb \cite{lhcb1,danielmstau2012}, $B(B^+ \to \pi^-\mu^+\mu^+) \leq 1.3 \times 10^{-8}$, the constraints on the $|V_{\mu N}|^2$ mixing are not largely improved owing to the Cabibbo suppresion factor described in the previous paragraph. 

\section{Four-body $B^0$ decays}
  
 In order to avoid the suppression due to CKM factors in the $B^+$ decay vertex, we have proposed  to consider the four-body decays of neutral mesons, namely $B^0 \to D^-M^-l^+l'^+$ with $l,\ l'=e,\ \mu$ or $\tau$ and $M$ a pseudoscalar or vector meson  \cite{nos2011}. The decay amplitude for this four-body decay is given by:
\begin{equation}
{\cal M}_4^{B^0} \sim G_F^2 V_{lN}V_{l'N}m_N{\cal F}(q^2)V_{cb}^{CKM}f_{M}F_+^{B\to D}(t)\ ,
\end{equation}
where $F_{+}^{B\to D}(t)$ is the vector form factor for the $B \to D$ transition and $t=(p_B-p_D)^2$ is the square of the momentum transfer (the contribution of the scalar form factor is negligible in this case). In this updated contribution we use the $B \to D$ vector form factor obtained from Lattice QCD \cite{kronfeld}, in order to avoid the model dependence of the vector form factor used in \cite{nos2011}.

  In the neutrino narrow width approximation, the generic expression for the branching ratios of three- and four-body decays can be written as:
\begin{equation}
B_{ll'}\sim \frac{|V_{lN}V_{l'N}|^2}{\Gamma_N}G(m_N)\ ,
\end{equation}
where $\Gamma_N=\sum_l f_l(m_N)|V_{lN}|^2$ is the neutrino decay width; the sum extends over the lepton flavors that are allowed by kinematics for a given neutrino mass $m_N$, and $f_l(m_N)$ depends on decay constants and masses of final states in neutrino decay channels \cite{atre}. The function $G(m_N)$ contains the product of fundamental constants, hadronic parameters as well as the integrated four-body phase space of the specific channels \cite{nos2011}. 

No upper limits have been reported so far for four-body LNV decays of neutral $B$ mesons. Upper limits for four-body LNV decays have been reported only in the case of $D^0\to M_1^-M_2^-l^+l'^+$ decays with $l,\ l'=e,\ \mu$ and $M_{1,2}=\pi,\ K$ \cite{e791}; in all cases, the upper limits obtained for the branching ratios are at the level of $10^{-5}\!\sim \!10^{-4}$ which loosely constrain the mixing angles. Very recently, the LHCb collaboration has reported the first upper limit on the charged $B$ decay channel, $B(B^-\to D^0\pi^+\mu^-\mu^-) \leq 1.5 \times 10^{-6}$ at the 95\% c. l. \cite{lhcb1}. The constraints in the $|V_{\mu N}|^2$ vs. $m_N$ plane that are obtained by assuming $B(B^0 \to D^-\mu^+\mu^+\pi^-) \leq 10^{-7}$ are shown in Figure 3. We observe that the constraint of the $\mu N$ mixing angle that can be obtained from this decay channel is competitive or even better when compared to other three-body LNV decays of $B^+$ mesons, for a mass region $m_N \leq 1.5$ GeV. 

\section{Four-body $\tau$ lepton decays}

  In order to look for better constraints on mixing of resonant neutrinos, we consider the $\tau^- \to \nu_{\tau}l^-l'^-M^+$ decays, where $l,l'=e$ or $\mu$ and $M=\pi, K, \rho$ and $K^*$ mesons. The use of these novel decay channels to constraint the parameter space of Majorana neutrinos were proposed in Ref.
\cite{nos2012}. A previous estimate of the branching ratio for the dimuonic channel $B(\tau^-\! \to \! \nu_{\tau}\mu^-\mu^-\pi^+) \leq 8.2 \times 10^{-5}$ was reported in \cite{chile2} using $|V_{\mu N}|^2 \leq 10^{-3}$ and $m_N= 400\sim 600$ MeV as typical values within a model where the sterile neutrino lifetime is dominated by the radiative $N\!\to\! \nu\gamma$ decay \cite{gninenko}. 

The four-body $\tau^- \to \nu_{\tau}l^-l'^-M^+$ decays can provide constrains directly on the $|V_{lN}|^2$ mixing angles contrary to their three-body decays which are sensitive only to the product $|V_{\tau N}V_{lN}|$. The decay amplitude in this case is given by 
\begin{equation}
{\cal M}_4^{\tau} \sim G_F^2V_{lN}V_{l'N}m_N{\cal F}(q^2)V^{CKM}_{uq}f_M\ ,
\end{equation}
where $q=d$ or $s$. 

\begin{figure}
\includegraphics[width=8.6cm]{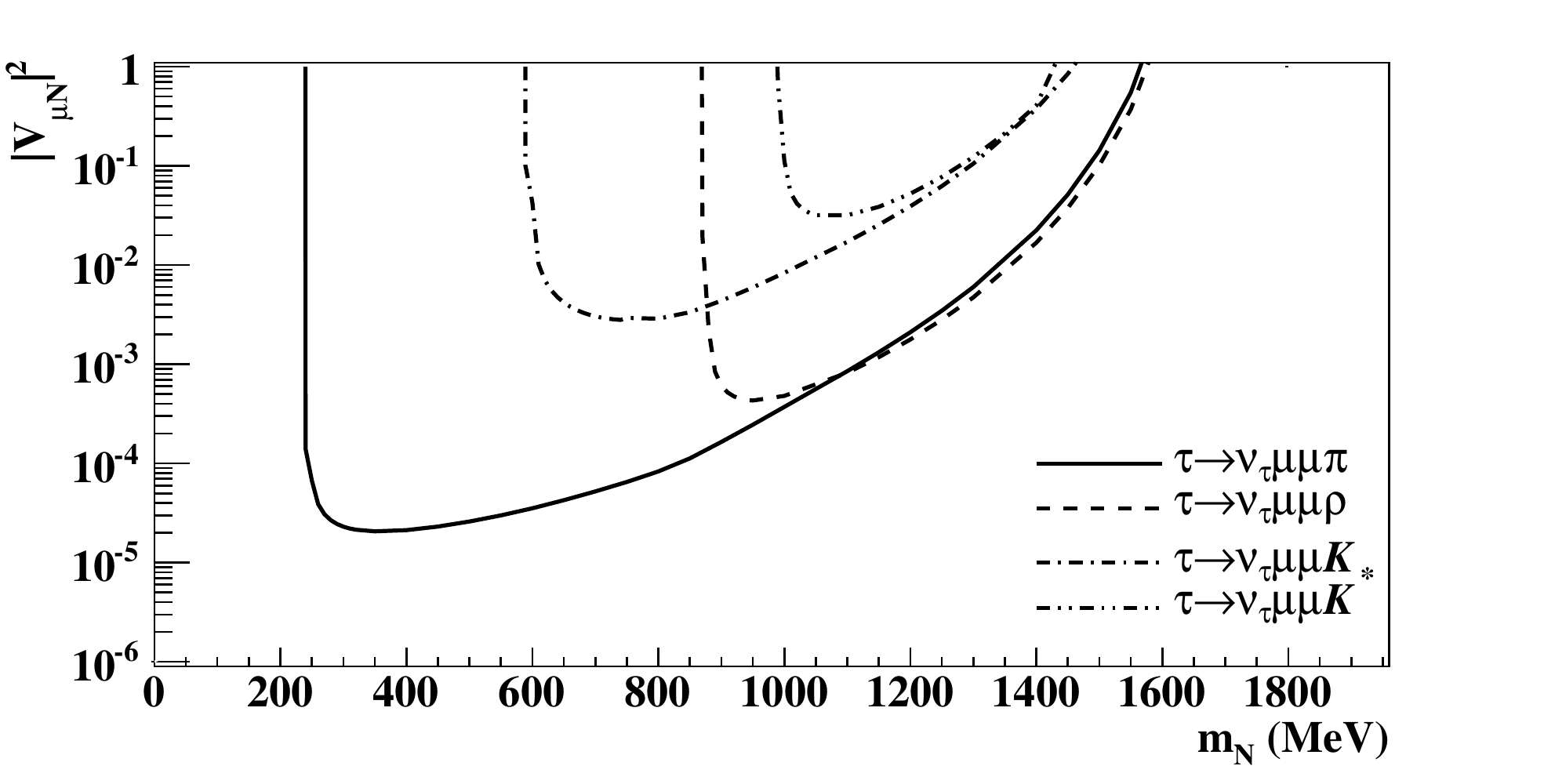}
\caption{\small Constraints on Majorana neutrino parameter space from four-body di-muonic channels of $\tau$ lepton decays.}
\end{figure}

 In Figure 4 we plot the constraints in the $|V_{\mu N}|^2$ vs. $m_N$ plane that are obtained by assuming a common upper limit $B(\tau^- \to \nu_{\tau}\mu^-\mu^-M^+) \leq 10^{-7}$ for all channels. In Figure 3, the constraints obtained by assuming $B(\tau^- \to \nu_{\tau}\mu^-\mu^-\pi^+)\leq 10^{-8}$ are compared to those obtained from heavy meson decays. As we can observe, these constraints are stronger than the ones that can be currently obtained from three-body decays of $B^+$ and $D^+$ mesons. In order to compare with the results of Ref. \cite{chile2}, we compute the following branching fraction (we use similar values of mixings and masses for the heavy sterile neutrino)
\begin{equation}
B(\tau^- \to \nu_{\tau}\mu^-\mu^-\pi^+) \leq 1.4 \times 10^{-5}\ .
\end{equation} 
Our results turns out to be of similar size. The main difference comes from the models we have used to compute the heavy neutrino lifetime.

  In summary, LNV decays of heavy flavors can provide important constraints on tiny mixing angles of Majorana neutrinos with masses in the range $m_{\pi} \leq m_N \leq m_B$. This is possible if a single heavy neutrino resonantly enhances the decay amplitudes. In this contribution we have shown that the four-body decays $B^0 \to D^-\pi^-\mu^+\mu^+$ and $\tau^- \to \nu_{\tau}\mu^-\mu^-\pi^+$ 
can provide stronger constrains on the $|V_{\mu N}|^2$ mixing angles than the ones obtained from three-body decays of charged heavy mesons.

  {\small The authors are grateful to the organizing committee of Tau2012 for the opportunity to present this work. They acknowledge the financial support from Conacyt (Mexico) and the collaboration of D. Delepine at the early stage of this work.}



%% References

\nocite{*}
\bibliographystyle{elsarticle-num}
\bibliography{martin}

\begin{thebibliography}{00}


\bibitem{moha}
R. N. Mohapatra {\it et al}, Rep. Prog. Phys. {\bf 70}, 1757 (2007).
\bibitem{paul}
P. G. Langacker, {\it The Standard Model and Beyond}, CRC Press Taylor and Francis Group, (2010).
\bibitem{rode}
For for example: S. M. Bilenky, Phys. Part. Nucl. {\bf 41}, 690 (2010); W. Rodejohann, Int. J. Mod. Phys. E{\bf 20}, 1833 (2011).
\bibitem{atre}
A. Atre, T. Han, S. Pascoli, and Zhang, JHEP 0905, 030 (2009).
\bibitem{ali}
A. Ali, A. V. Borisov and N. B. Zamorin, Eur. Phys. J. C{\bf 21}, 123 (2001).
\bibitem{atre1}
A. Atre, V. Barger and T. Han, Phys. Rev. D{\bf 79}, 113014 (2005)
\bibitem{chile1}
J. C. Helo, S. Kovalenko and I. Schmidt, Nucl. Phys. B853, 80 (2011).
\bibitem{bao}
M. A. Ivanov and S. G. Kovalenko, Phyd. Rev. D{\bf 71}, 053004 (2005); S. S. Bao et al, arXiv:1208.5136 [hep-ph]; J. M. Chang and G. L. Wang, Eur. Phys. C{\bf 71}, 1715 (2011).
\bibitem{babar1}
J. P. Lees et al [BABAR Collab.], Phys. Rev. D{\bf 84}, 072006 (2011); Phys. Rev. D{\bf 85}, 071103(R) (2012).
\bibitem{lhcb1}
R. Aaij et al [LHCb Collab.], Phys. Rev. Lett. {\bf 108}, 101601 (2012); Phys. Rev. D{\bf 85}, 112004 (2012).
\bibitem{belle2}
O. Seon {\it et al} [Belle Collab.], Phys. Rev. D{\bf 84}, 071106(R) (2011).
\bibitem{belle1}
Y. Miyazaki et al [Belle Collab.], Phys. Lett. B682, (2010).
\bibitem{pdg}
J. Beringer {\it et al}, Phys. Rev. D{\bf 86}, 010001 (2012).
\bibitem{danielmstau2012}
D. Mart\'\i nez Santos, {\it these proceedings}
\bibitem{kiyoshitau2012}
K. Hasayaka, {\it these proceedings}
\bibitem{shapo}
L. Canneti {\it et al}, arXiv:1208.4607 [hep-ph]. 
 \bibitem{nos2011}
D. Delepine, G. L\'opez Castro and N. Quintero, Phys. Rev. D{\bf 84}, 096011 (2011); {\it ibid} D{\bf 86}, 079905(E) (2012).
\bibitem{nos2012}
G. L\'opez Castro and N. Quintero, Phys. Rev. D{\bf 85}, 076006 (2012); {\it ibid} D{\bf 86}, 079904(E) (2012).
\bibitem{chile3}
G. Cvetic, C. Dib, and C. S. Kim, JHEP1206, 149 (2012). 
\bibitem{e791}
E. M. Aitala {\it et al} [E791 Collab.], Phys. Rev. Lett. 86, (2001).
\bibitem{kronfeld}
J. A. Bailey {\it et al} [Fermilab Lattice and MILC], Phys. Rev. D{\bf 85}, 114502 (2012); {\it ibid} D{\bf 86}, 039904(E) (2012).
\bibitem{chile2}
C. Dib et al, Phys. Rev. D{\it 85}, 011301(R) (2012).
\bibitem{gninenko}
S. N. Gninenko, Phys. Rev. Lett. {\bf 103}, 241802 (2009).
\end{thebibliography}



\end{document}